\documentclass[prX,showpacs,nofootinbib]{revtex4}
\usepackage{amssymb}  

\newcommand{\be}{\begin{equation}}
\newcommand{\ee}{\end{equation}}
\newcommand{\ba}{\begin{eqnarray}}
\newcommand{\ea}{\end{eqnarray}}

\def\a{\alpha}
\def\b{\beta}
\def\d{\delta}
\def\e{\epsilon}

\def\f{\phi}
\def\vf{\varphi}

\def\m{\mu}
\def\n{\nu}

\def\p{\pi}

\def\r{\rho}

\def\t{\tau}

\def\D{\Delta}
\def\F{\Phi}
\def\G{\Gamma}

\def\P{\Pi}

\def\S{\Sigma}

\def\cg{{\cal G}}

\def\cs{{\cal S}}

\def\cw{{\cal W}}

\newcommand{\ov}{\overline}

\newcommand{\wt}{\widetilde}

\newcommand{\aand}{\;\;\;\mbox{and}\;\;\;}
\newcommand{\pa}{\partial}

\def\bc{\bar c}
\def\bx{\bar\xi}

\def\I{\leavevmode\hbox{\small1\kern-3.8pt\normalsize1}}

\begin{document}

\title{The Jackiw-Pi model and its symmetries}
\author{O.M. Del Cima}
\email{oswaldo.delcima@ufv.br}
\affiliation{Universidade Federal de Vi\c cosa (UFV),\\
Departamento de F\'\i sica - Campus Universit\'ario,\\
Avenida Peter Henry Rolfs s/n - 36570-000 -
Vi\c cosa - MG - Brazil.}
\date{\today}

\begin{abstract}
The non-Abelian gauge model proposed by Jackiw and Pi, 
which generates an even-parity mass term in three 
space-time dimensions, is revisited in this letter. 
All the symmetries of the 
model are collected and established by means of BRS invariance and Slavnov-Taylor identity. 
The path for the 
perturbatively quantization of the Jackiw-Pi model, through the algebraic method of renormalization, is presented.
  
\end{abstract}
\maketitle


One of the central problems in the framework of gauge field theories is the issue of gauge field mass. Gauge
symmetry is not, in principle, conflicting with the presence of a massive gauge boson. In two space-time
dimensions, the well-known Schwinger model puts in evidence the presence of a massive photon without the
breaking of gauge symmetry~\cite{schwinger}. Another evidence for the compatibility between gauge symmetry and massive
vector fields has been arisen in the study of three-dimensional gauge theories, when a topological mass term referred to as
the Chern-Simons one, once added to the Yang-Mills term, shifts the photon mass to a non-vanishing
value without breaking gauge invariance, however parity symmetry is lost~\cite{deser-jackiw-templeton}. 
In 1997, Jackiw and Pi overcame the challenge to implement both gauge and parity invariance in three space-time 
dimensions by breaking the Yang-Mills paradigm - non-Abelian generalizations of Abelian models. They proposed a  
three-dimensional non-Yang-Mills gauge model for a pair of vector fields with opposite parity 
transformations, which generates a mass-gap through a mixed Chern-Simons-like term preserving parity~\cite{jackiw-pi}. 
Later on, the Jackiw-Pi model has been studied in the Hamiltonian framework, where physical states consistency 
was demonstrated~\cite{dayi}.    
In this letter, the non-Abelian gauge model proposed by Jackiw and Pi, which generates
an even-parity mass term in three space-time dimensions, is revisited. The symmetries of the model are collected and
established through BRS invariance and Slavnov-Taylor identity, also the BRS approach has allowed to bypass the difficulties 
cited in the literature with respect to the gauge-fixing. In the Landau gauge, thanks to the antighost
equations and the Slavnov-Taylor identity, two rigid symmetries are identified by means of Ward identities. 
The propagators computation, the spectrum consistency and the tree-level unitarity analysis are left to be 
presented in a forthcoming paper~\cite{jackiwpiclassical}. The 
Jackiw-Pi model remains unquantized up to now, however, it is presented here the key ingredients for its further 
perturbatively quantization through the algebraic method of renormalization~\cite{jackiwpiquantum}.

\section{The model and its symetries}

\subsection{The model}

The classical action of the Jackiw-Pi model~\cite{jackiw-pi} is given by:
\be
\S_{\rm inv}={\rm Tr}\int{d^3 x} \left\{
{1\over2}F^{\m\n}F_{\m\n} 
+ {1\over2}\bigl(G^{\m\n}+g[F^{\m\n},\r]\bigr)\bigl(G_{\m\n}+g[F_{\m\n},\r]\bigr) 
- m\e^{\m\n\r}F_{\m\n} \f_\r \right\}~,
\label{action}
\ee
such that,
\be
F_{\m\n}=\pa_\m A_\n - \pa_\n  A_\m + g[A_\m,A_\n]~,~~G_{\m\n}=D_\m \f_\n - D_\n \f_\m 
\aand D_\m\bullet=\pa_\m\bullet + g[A_\m,\bullet~]~,
\ee
where $A_\m$ and $\f_\m$ are vector fields, $\r$ is a scalar, $g$ is a coupling constant and $m$ a mass parameter, also, $\bullet$ means any field. Every field, $X=X_a\t_a$, is Lie algebra valued, where the matrices $\t$ are the generators of the group and obey 
\be
[\t_a,\t_b]=f_{abc}\t_c \aand {\rm Tr}(\t_a\t_b)=-{1\over2}\d_{ab}~.
\ee
 
\subsection{Gauge symmetries} 

The action (\ref{action}) is invariant under two sets of gauge transformations, $\d_\theta$ and $\d_\chi$:
\ba
&&\d_\theta A_\m=D_\m\theta ~,~~ \d_\theta \f_\m=g[\f_\m,\theta] \aand \d_\theta \r=g[\r,\theta] ~;\label{theta}\\
&&\d_\chi A_\m=0 ~,~~ \d_\chi \f_\m=D_\m\chi  \aand \d_\chi \r=-\chi ~,\label{chi}
\ea
where $\theta$ and $\chi$ are Lie algebra valued infinitesimal local parameters.

\subsection{BRS symmetry} 

The corresponding BRS transformations of the fields $A_\m$, $\f_\m$ and $\r$, stemming from the symmetries 
(\ref{theta}) and (\ref{chi}), are given by\footnote{The commutators among the fields are assumed to 
be graded, namely, $[\vf_1^{g_1},\vf_2^{g_2}]
\equiv\vf_1^{g_1}\vf_2^{g_2}-(-1)^{g_1.g_2}\vf_2^{g_2}\vf_1^{g_1}$, where the 
upper indices, $g_1$ and $g_2$, are the Faddeev-Popov charges ($\F\P$) carried 
by the fields $\vf_1^{g_1}$ and $\vf_2^{g_2}$, respectively.}:
\ba
&& sA_\m=D_\m c ~,~~ s\f_\m=D_\m \xi + g[\f_\m,c] ~,~~ s\r=-\xi + g[\r,c]~, \nonumber\\
&& sc=-gc^2 \aand s\xi=-g[\xi,c]~, \label{BRS}
\ea
where $c$ and $\xi$ are the Faddeev-Popov ghosts, with Faddeev-Popov charge 
(ghost number) one. The ghost number ($\F\P$) of all 
fields and antifields are collected in Table~\ref{dimensions}

\subsection{The gauge-fixing and the antifields action} 

The gauge-fixing adopted here belongs to the class of the linear covariant gauges discussed 
by 't Hooft~\cite{thooft}. In order to implement the gauge-fixing following the BRS procedure~\cite{brs,piguet}, we introduce two sorts of ghosts ($c$ and $\xi$), antighosts ($\bc$ and $\bx$) and the Lautrup-Nakanishi fields~\cite{lautrup-nakanishi} ($b$ and $\p$), playing the role of Lagrange multiplier fields 
for the gauge condition, such that  
\ba
&&s\bc=b~,~~sb=0~;  \label{doubletbc}\\
&&s\bx=\p~,~~s\p=0~;\label{doubletbd}
\ea
where the multiplier fields, $b$ and $\p$, and the Faddeev-Popov antighosts, $\bc$ and $\bx$, with ghost number minus one, belong to the BRS-doublets (\ref{doubletbc}) and (\ref{doubletbd}).

Now, by adopting the gauge conditions
\ba
\frac{\d \S_{\rm{gf}}}{\d b}= \pa^\m A_\m + \a b~,\label{gaugefixing1}\\
\frac{\d \S_{\rm{gf}}}{\d \p}= \pa^\m \f_\m + \b \p~,\label{gaugefixing2}
\ea
it follows that the BRS-trivial gauge-fixing action compatible with then reads
\ba
\S_{\rm{gf}}&\!\!=\!\!&s~{\mbox{Tr}}\int d^3x~\left\{ \bc\pa^\m A_\m + \bx
\pa^\m \f_\m\ + \frac{\a}{2} \bc b + \frac{\b}{2}\bx \p \right\} \nonumber\\
&\!\!=\!\!&{\mbox{Tr}}\int d^3x~\left\{ b\pa^\m A_\m - \bc \pa^\m D_\m c 
+ \p \pa^\m \f_\m\ - \bx \pa^\m \bigl( D_\m \xi + g[\f_\m,c] \bigr) 
+ \frac{\a}{2} b^2 + \frac{\b}{2} \p^2 \right\}~.
\ea
Let us now introduce the action in which the nonlinear BRS transformations are coupled to the the antifields 
(BRS invariant external fields), so as to control, at the quantum level, the (further) renormalization of those transformations:
\be
\S_{\rm{ext}}={\mbox{Tr}}\int d^3x~\left\{ A^*_\m sA^\m + \f^*_\m s\f^\m 
+ \r^* s\r + c^* sc + \xi^* s\xi \right\}~,
\ee
where, as mentioned above, the antifields are BRS invariant, namely, 
\be
sA^*_\m=s\f^*_\m=s\r^*=sc^*=s\xi^*=0~.\label{BRSanti}
\ee

The total action at the tree level for the Jackiw-Pi model, $\G^{(0)}$, 
is therefore given by:
\be
\G^{(0)}=\S_{\rm inv} + \S_{\rm{gf}} + \S_{\rm{ext}}~,\label{totalaction}
\ee
which is invariant under the BRS transformations given by the equations (\ref{BRS}), (\ref{doubletbc}), (\ref{doubletbd}) and (\ref{BRSanti}). The action (\ref{totalaction}) preserves the ghost number. The values of the ghost number, the ultraviolet (UV) and the infrared (IR) dimensions (respected 
to the Landau gauge) are displayed in Table~\ref{dimensions}. However, all subtleties concerning the determination of the UV and the IR dimensions of the fields, in the Landau gauge, shall be presented in Ref.\cite{jackiwpiclassical}. The statistics is defined as follows: the fields of integer spin and odd ghost number as well as the fields of half integer spin and even ghost number are 
anticommuting; the other fields commute with the formers and among themselves.

An interesting feature of the Jackiw-Pi action $\G^{(0)}$(\ref{totalaction}) is that it is not BRS local 
invariant thanks to the parity-even mass term:
\be
\S_{\rm m}={\rm Tr}\int{d^3 x} \left\{- m\e^{\m\n\r}F_{\m\n} \f_\r \right\}~,
\ee 
since
\be
sF_{\m\n}=g[F_{\m\n},c]~,
\ee
then
\be
s\S_{\rm m}=-m~s{\rm Tr}\int{d^3 x} \left\{\e^{\m\n\r}F_{\m\n} \f_\r \right\}= 
-m~{\rm Tr}\int{d^3 x} \left\{ \e^{\r\m\n}\pa_\r(F_{\m\n}\xi) \right\}~,\label{nonbrslocalinv}
\ee
which is invariant only up to a total derivative, possibly indicating that at the quantum level the $\b$-function 
associated to the mass parameter $m$ vanishes~\cite{delcima-franco-helayel-piguet,barnich}.

\subsection{Slavnov-Taylor identity, ghost and antighost equations and Ward identities} 

This subsection is devoted to establish the Slavnov-Taylor identity, ghost and antighost equations, and two 
hidden rigid symmetries.
The BRS invariance of the action $\G^{(0)}$ (\ref{totalaction}) is expressed through the Slavnov-Taylor 
identity
\ba
\cs(\G^{(0)})&=&{\mbox{Tr}}\int d^3x~\Bigg\{ \frac{\d\G^{(0)}}{\d A^*_\m} \frac{\d\G^{(0)}}{\d A^\m}    
+ \frac{\d\G^{(0)}}{\d \f^*_\m} \frac{\d\G^{(0)}}{\d \f^\m} 
+ \frac{\d\G^{(0)}}{\d \r^*} \frac{\d\G^{(0)}}{\d \r}
+ \frac{\d\G^{(0)}}{\d c^*} \frac{\d\G^{(0)}}{\d c}
+ \frac{\d\G^{(0)}}{\d \xi^*} \frac{\d\G^{(0)}}{\d \xi} \nonumber\\
&+& b \frac{\d\G^{(0)}}{\d \bc} + \p \frac{\d\G^{(0)}}{\d \bx}\Bigg\}=0~,\label{Slavnov}
\ea
which translates, in a functional way, the invariance of the classical model under the BRS symmetry. It is 
suitable to define, for later use, the linearized Slavnov-Taylor ($\cs_{\G^{(0)}}$) operator as below
\ba
\cs_{\G^{(0)}}&=&{\mbox{Tr}}\int d^3x~\Bigg\{ \frac{\d\G^{(0)}}{\d A^*_\m} \frac{\d}{\d A^\m}    
+ \frac{\d\G^{(0)}}{\d A^\m} \frac{\d}{\d A^*_\m} 
+ \frac{\d\G^{(0)}}{\d \f^*_\m} \frac{\d}{\d \f^\m} 
+ \frac{\d\G^{(0)}}{\d \f^\m} \frac{\d}{\d \f^*_\m}
+ \frac{\d\G^{(0)}}{\d \r^*} \frac{\d}{\d \r}
+ \frac{\d\G^{(0)}}{\d \r} \frac{\d}{\d \r^*} \nonumber\\
&+& \frac{\d\G^{(0)}}{\d c^*} \frac{\d}{\d c}
+ \frac{\d\G^{(0)}}{\d c} \frac{\d}{\d c^*}
+ \frac{\d\G^{(0)}}{\d \xi^*} \frac{\d}{\d \xi} 
+ \frac{\d\G^{(0)}}{\d \xi} \frac{\d}{\d \xi^*}
+ b \frac{\d}{\d \bc} + \p \frac{\d}{\d \bx}\Bigg\}~.
\ea
Another identities, the ghost equations
\ba
\cg_{\rm I}\G^{(0)}\equiv\frac{\d\G^{(0)}}{\d \bc}
+\pa^\m\frac{\d\G^{(0)}}{\d A^{*\m}}=0~,\\
\cg_{\rm II}\G^{(0)}\equiv\frac{\d\G^{(0)}}{\d \bx}
+\pa^\m\frac{\d\G^{(0)}}{\d \f^{*\m}}=0~,
\ea
follow from the gauge-fixing conditions, (\ref{gaugefixing1}) and (\ref{gaugefixing2}), and from Slavnov-Taylor identity (\ref{Slavnov}), 
meaning that $\G^{(0)}$ depends on the antighosts, $\bc$ and $\bx$, and the antifields, $A^{*\m}$ and $\f^{*\m}$, 
through the combinations 
\be
{\wt A}^*_\m=A^*_\m+\pa_\m\bc \aand 
{\wt \f}^*_\m=\f^*_\m+\pa_\m\bx~.
\ee

The Jackiw-Pi model presents two antighost equations, they are listed as below:
\ba
\ov\cg_{\rm I}\G^{(0)}&\equiv&\int d^3x~\biggl\{\frac{\d\G^{(0)}}{\d c}
-g\biggl[\bc,\frac{\d\G^{(0)}}{\d b}\biggr]-g\biggl[\bx,\frac{\d\G^{(0)}}{\d \p}\biggr]\biggr\}={\ov\D}_{\rm I}~, \label{antighost1} \\
{\mbox{where}}~~{\ov\D}_{\rm I}&\equiv&-g\int d^3x~\bigl\{[A^*_\m,A^\m]+[\f^*_\m,\f^\m]+[\r^*,\r]-[c^*,c]-[\xi^*,\xi]
+ \a[\bc,b]+\b[\bx,\p]\bigr\}~;\label{delta1}\\
\ov\cg_{\rm II}\G^{(0)}&\equiv&\int d^3x~\biggl\{\frac{\d\G^{(0)}}{\d \xi}
-g\biggl[\bx,\frac{\d\G^{(0)}}{\d b}\biggr]\biggr\}={\ov\D}_{\rm II}~,\label{antighost2}\\
{\mbox{where}}~~{\ov\D}_{\rm II}&\equiv&-g\int d^3x~\bigl\{[\f^*_\m,A^\m]-[\xi^*,c]-\frac{\r^*}{g}
+ \a[\bx,b]\bigr\}~.\label{delta2}
\ea
It should be noticed, for the sake of later quantization~\cite{jackiwpiquantum}, that the breakings, ${\ov\D}_{\rm I}$ and ${\ov\D}_{\rm II}$, being nonlinear in the quantum fields will be subjected to renormalization. An interesting issue in Yang-Mills theories is that the Landau gauge~\cite{landaugauge} has very special features as compared to a generic linear gauge. This is due to the existence, besides the Slavnov-Taylor identity, of 
another identity, the antighost equation~\cite{blasi-piguet-sorella}, which controls the dependence of the theory on the ghost $c$. In particular, 
this equation implies that the ghost field $c$ and the composite $c$-field cocycles in the 
descent equations have vanishing anomalous dimension, allowing the algebraic proof~\cite{piguet} of the Adler-Bardeen 
nonrenormalization theorem~\cite{adler-bardeen} for the gauge anomaly. Back to the Jackiw-Pi model we are considering here, in the case of the general linear covariant gauges, (\ref{gaugefixing1}) and (\ref{gaugefixing2}), the right-hand sides of the equations, (\ref{antighost1}) and 
(\ref{antighost2}), are nonlinear in the quantum fields due to the presence of the terms, 
$\int d^3x~\a[\bc,b]$ and $\int d^3x~\b[\bx,\p]$, and $\int d^3x~\a[\bx,b]$, respectively. Therefore, 
the breakings, ${\ov\D}_{\rm I}$ (\ref{delta1}) and ${\ov\D}_{\rm II}$ (\ref{delta2}), in the quantized theory, have to be renormalized, 
which could spoil the usefulness of the antighost equations, by this reason, bearing in mind the (further) 
renormalization of the model~\cite{jackiwpiquantum}, we adopt from now on the Landau gauge $\a=\b=0$. 

As another feature of the Landau gauge, the following Ward identities for the rigid symmetries stem from the Slavnov-Taylor identity 
(\ref{Slavnov}) and the antighost equations (\ref{antighost1}) and (\ref{antighost2}) with $\a=\b=0$:
\ba
\cw_{\rm I}^{\rm rig}\G^{(0)}&=&0~,~~
{\mbox{where}} \label{rigid1} \nonumber\\
\cw_{\rm I}^{\rm rig}&\equiv& -g \int d^3x~\biggl\{ \biggl[A^\m,\frac{\d}{\d A^\m}\biggr] + 
\biggl[\f^\m,\frac{\d}{\d \f^\m}\biggr] + \biggl[\r,\frac{\d}{\d \r}\biggr] + 
\biggl[b,\frac{\d}{\d b}\biggr] + \biggl[\p,\frac{\d}{\d \p}\biggr] + \biggl[c,\frac{\d}{\d c}\biggr] + 
\biggl[\xi,\frac{\d}{\d \xi}\biggr] + \nonumber\\
&+& \biggl[\bc,\frac{\d}{\d \bc}\biggr] + \biggl[\bx,\frac{\d}{\d \bx}\biggr] + 
\biggl[A^*_\m,\frac{\d}{\d A^*_\m}\biggr] + \biggl[\f^*_\m,\frac{\d}{\d \f^*_\m}\biggr] + 
\biggl[\r^*,\frac{\d}{\d \r^*}\biggr] + \biggl[c^*,\frac{\d}{\d c^*}\biggr] + \biggl[\xi^*,\frac{\d}{\d \xi^*}\biggr]
\biggr\}~;\\
\cw_{\rm II}^{\rm rig}\G^{(0)}&=&0~,~~
{\mbox{where}} \label{rigid2} \nonumber\\
\cw_{\rm II}^{\rm rig}&\equiv& -g\int d^3x~\biggl\{ \biggl[A^\m,\frac{\d}{\d \f^\m}\biggr] + 
\biggl[\p,\frac{\d}{\d b}\biggr] + \biggl[c,\frac{\d}{\d \xi}\biggr] + \biggl[\bx,\frac{\d}{\d \bc}\biggr] + 
\biggl[\f^*_\m,\frac{\d}{\d A^*_\m}\biggr] + \biggl[\xi^*,\frac{\d}{\d c^*}\biggr] + \frac{1}{g}\frac{\d}{\d \r} \biggr\}~,
\ea



\begin{table}[t]
\begin{center}
\begin{tabular}{|c||c|c|c|c|c|c|c|c|c|c|c|c|c|c|c|c|}
\hline
& $A_\m$ & $\f_\m$ & $\r$ & $b$ & $\p$ & $c$ & $\xi$ & ${\bc}$ & ${\bx}$ 
& $A^{*\m}$ & $\f^{*\m}$ & $\r^*$ & $c^*$ & $\xi^*$ & $g$ & $m$\\ 
\hline\hline
$d$ & $1/2$ & $1/2$ & $-1/2$ & $3/2$ & $3/2$ & $-1/2$ & $-1/2$ & $3/2$ & $3/2$ & $5/2$ & $5/2$ & $7/2$ & $7/2$ & $7/2$ & $1/2$ & $1$\\ \hline
$r$ & $1/2$ & $1/2$ & $-1/2$ & $3/2$ & $3/2$ & $-1/2$ & $-1/2$ & $3/2$ & $3/2$ & $5/2$ & $5/2$ & $7/2$ & $7/2$ & $7/2$ & $1/2$ & $1$\\ \hline
$\Phi\Pi$ & $0$ & $0$ & $0$ & $0$ & $0$ & $1$ & $1$ & $-1$ & $-1$ & $-1$ & $-1$ & $-1$ & $-2$ & $-2$ & $0$ & $0$\\ \hline
\end{tabular}
\end{center}
\caption[]{Ultraviolet dimension ($d$), infrared dimension ($r$) and ghost number ($\Phi \Pi$)\label{dimensions}.}
\end{table}

\section{Conclusions}

The Jackiw-Pi model~\cite{jackiw-pi} which generates a mass gap preserving parity in three space-time dimensions was presented here. 
The BRS symmetry of the model was established and the difficulties cited in the literature concerning the gauge-fixing 
were bypassed. Also, BRS invariance and Slavnov-Taylor identity together with the antighost equations, in the Landau gauge, have allowed to 
find out two rigid symmetries. In spite of being out of the scope of this letter, an important issue to be noticed is that, as we have shown 
in (\ref{nonbrslocalinv}), the Jackiw-Pi even-parity mass term is BRS 
invariant up to a total derivative, {\it i.e.}, it is not local BRS invariant. Therefore, it could be conjectured that, at the quantum level, the $\b$-function associated to the mass parameter $m$, $\b_m$, should be zero, $\b_m=0$~\cite{delcima-franco-helayel-piguet, barnich}. Moreover, this fact would indicate the perturbatively ultraviolet finiteness of the Jackiw-Pi model, which is now under investigation~\cite{jackiwpiquantum} in the framework of the algebraic renormalization scheme.

\subsection*{Acknowledgements}
O.M.D.C. dedicates this work to his kids, Vittoria and Enzo, and to his mother, Victoria.



\begin{references}

\bibitem{schwinger} J. Schwinger, Phys. Rev. 125 (1962) 397; Phys. Rev. 128 (1962) 2425.

\bibitem{deser-jackiw-templeton} R. Jackiw and S. Templeton, Phys. Rev. D23 (1981) 2291; 
J. Schonfeld, Nucl. Phys. B185 (1981) 157; S. Deser, R. Jackiw and S. Templeton, 
Ann. Phys. (NY) 140 (1982) 372; Phys. Rev. Lett. 48 (1982) 975.  

\bibitem{jackiw-pi} R. Jackiw and S.-Y. Pi, Phys. Lett. B403 (1997) 297; 
R. Jackiw, {\it Non-Yang-Mills gauge theories}, hep-th/9705028, 
lectures given at Advanced Summer School on Non-perturbative Quantum 
Field Physics, Peniscola, Spain, June 1997. 

\bibitem{dayi} \"O.F. Dayi, Mod. Phys. Lett. A13 (1998) 1969.

\bibitem{jackiwpiclassical} O.M. Del Cima, {\it The Jackiw-Pi model I: classical theory}, 
work in progress.

\bibitem{jackiwpiquantum} O.M. Del Cima, D.H.T. Franco and O. Piguet, {\it The Jackiw-Pi model II: quantum theory}, 
work in progress. 

\bibitem{thooft} G. 't Hooft, Nucl. Phys. B33 (1971) 173; Nucl. Phys. B35 (1971) 167.

\bibitem{brs} C. Becchi, A. Rouet and R. Stora, Ann. Phys. (N.Y.) 98 (1976) 287; 
T. Kugo and I. Ojima, Phys. Lett. B73 (1978) 459; Progr. Theor. Phys. Suppl. 66 (1979) 1.

\bibitem{piguet} O. Piguet and S.P. Sorella, \textit{Algebraic Renormalization}, 
Lecture Notes in Physics, m28, Springer-Verlag (Berlin-Heidelberg), 1995; 
see also references therein.

\bibitem{lautrup-nakanishi} B. Lautrup, Mat. Fys. Medd. Dan. Vid. Selsk 35 (1967) No.11; 
N. Nakanishi, Progr. Theor. Phys. 35 (1966) 1111; Progr. Theor. Phys. 37 (1967) 618.

\bibitem{delcima-franco-helayel-piguet} O.M. Del Cima, D.H.T. Franco, J.A. Helay\"el-Neto and O. Piguet; 
 JHEP 9802 (1998) 002; JHEP 9804 (1998) 010; Lett. Math. Phys. 47 (1999) 265. 

\bibitem{barnich} G. Barnich, JHEP 12 (1998) 003.


\bibitem{landaugauge} G. Leibbrandt, Rev. Mod. Phys. 59 (1987) 1067.

\bibitem{blasi-piguet-sorella} A. Blasi, O. Piguet and S.P. Sorella, Nucl. Phys. B356 (1991) 154.

\bibitem{adler-bardeen} S.L. Adler and W.A. Bardeen, Phys. Rev. 182 (1969) 1517.






\end{references}
\end{document}